\documentclass[onecollarge,natbib]{svjour2}
\bibpunct{[}{]}{,}{n}{}{,} 
\smartqed  
\usepackage{graphicx}
\usepackage{booktabs}
\usepackage{multirow}
\journalname{Few-Body Systems}
\begin{document}

\title{Charmonium states in QCD-inspired quark potential model using Gaussian expansion method}

\author{Lu Cao         \and
        You-Chang Yang \and
        Hong Chen
}


\institute{Lu Cao \at
              School of Physical Science and Technology, Southwest University, Chongqing 400700, China \\
              \email{physicaolu@126.com}
             \and
           You-Chang Yang \at
              Department of Physics, Zunyi Normal College, Zunyi 563002, China\\
               \email{yangyouchang@yahoo.com.cn}
              \and
           Hong Chen\at
              School of Physical Science and Technology, Southwest University, Chongqing 400700, China \\
              Corresponding Author.
          \email{chenh@swu.edu.cn}
}

\date{Received: date / Accepted: date}

\maketitle

\begin{abstract}
We study the mass spectrum and electromagnetic processes of charmonium system with the spin-dependent potentials fully taking into account in the solution of the Schroedinger equation and the results for the pure scalar and scalar-vector mixing linear confining potentials are compared. It is revealed that the scalar-vector mixing confinement is important for reproducing the mass spectrum and decay widths and the vector component is found to
be around $22 \% $, the long-standing discrepancy in M1 radiative transitions of $J/\psi$ and $\psi^{\prime}$ is alleviated by means of the state wave functions obtained via the Hamiltonian with the full spin-dependent potential. This work also intends to identify few of the copious higher charmonium-like states as the $c\bar{c}$ ones. Particularly, the newly observed $X(4160)$ and $X(4350)$ are assigned as $M(2^1D_2)= 4164.9$ MeV and $M(3^3P_2)= 4352.4$ MeV, which strongly favor the $J^{PC}=2^{-+}, 2^{++}$ assignments respectively. The corresponding radiative transitions, leptonic and two-photon decay widths have been also calculated for the further experimental study.

\keywords{charmonium \and potential model \and Gaussian expansion method \and XYZ mesons }
\end{abstract}

\section{Introduction}
\label{intro}
Due to the impressive increase of experimental results, charmonium ($c\bar{c}$) spectroscopy has renewed great interest recently, coming along with the striking disagreement with theoretical expectations \cite{puzzles,babar2010,recharm}. The unexpected and still-fascinating $X(3872)$ has been joined by more than a dozen other charmonium-like states, while the series of vacancy have been left on the $c\bar{c}$ list. It is urgent to identify the possible
new members of charmonium family from the abundant observations.

The QCD inspired potential models have been playing an important role in investigating heavy quarkonium, owning to the presence of large nonperturbative effects in this energy region. Most quark potential models \cite{QWG169,prd17,prd21,prd32,lhqq,DEbert02,Bc,sop-heavy,prd60-Bc,prd50,prd49hqp,prd53-Bc,mesonRelat} have common ingredients under the non-relativistic limit, despite some differences in the detailed corrections for relativistic and coupled channel effects, which typically are the Coulomb-like term induced by one-gluon exchange plus the long-range confining potential expected from nonperturbative QCD. Anyway, the nature of confining mechanism has been veiled so far. In the original Cornell model \cite{cornell01,cornell02}, it was assumed as Lorentz scalar, which gives a vanishing long-range magnetic contribution and agrees with the flux tube picture of quark confinement \cite{flux}. Another possibility \cite{DEbert02,DEbert03} is that confinement may be a more complicated mixture of scalar and timelike vector, while the vector potential is anticonfinning. In pure $c\bar{c}$ models, the Lorentz nature of confinement is tested by the multiplet splitting of orbitally excited charmonium states.

In addition, numerical precision of the calculation method is important in testing different models. As several numerical methods fail in the potentials with $1/r^2$ and even higher negative power, the $\mathcal {O}(v^2/c^2)$ corrections to the quark-antiquark potential has to be usually treated as perturbation \cite{prd72,screen79,DEbert03}. However, the accuracy of perturbation expansion has been alerted recently \cite{prd75}, which indicates the most significant effect of the different treatments is on the wave functions. The exact solution of the full Hamiltonian provides every state with its own wave function, while the perturbative treatment leads to the same angular momentum multiplets sharing the identical radial wave function. It is known that the radiative transitions, leptonic and double-photon decay widths are quite sensitive to the shape of wave function and its behavior at
the origin. Each physical particle with different quantum numbers should have different state wave function. Thus, it is interesting to study the still-puzzling confining mechanism and the numerical precision of the calculation method.

In our calculation framework, the spin-dependent and -independent interactions have been totally taken into account in the Hamiltonian, where the different confining assumptions are compared from the mass spectrum and decay properties. The precise wave functions are inspected through the electromagnetic processes, i.e. radiative transitions and leptonic decays, which are considered to be a nichetargeting test for the overlap of radial integration and the subtle information at the origin. Since the relativistic reconstruction of the static confining potential is not unique, it
complicates the nature of confinement. Hence special concerns are focused on the minimal but relatively well-understood models, with the aim of gleaning the actual influences of different forms of confinement potential and the difference between the exact and the perturbative solutions.

In the following section, the two potential models are described in detail, along with the adopted variational approach and the optimization of parameters. The numerical results are assembled in Sec.\ref{sec:3}, a discussion related to the latest experimental results are put in Sec.\ref{sec:4}. Finally, Sec.\ref{sec:5} summaries the remarks and conclusions.

\section{Potential models and calculational approach}
\label{sec:2}

The confinement of quarks is assumed to be purely scalar linear type in NR model\cite{prd72}, and the scalar-vector mixed one in MNR \cite{DEbert02}.Once the Lorentz structure of central part are fixed in the two forms,
\begin{eqnarray}
\mathrm{NR}:\;  &&    V_S= b\,r ; \;\; V_V=-\frac{4}{3}\frac{\alpha _s}{r}, \\
\mathrm{MNR}: \;  &&  V_S= b(1-\epsilon)r; \;\; V_V=-\frac{4}{3}\frac{\alpha _s}{r}+\epsilon b r, \end{eqnarray}
the spin-orbit term and the tensor term can be directly derived from the standard Breit-Fermi expression to order $(v^2/c^2)$ with the charmed quark mass $m_c$. Summarily, the interaction potentials are

\begin{equation}
V_{NR}= -\frac{4}{3}\frac{\alpha _s}{r}+b\,r+\frac{32\pi \alpha _s}{9m_c^2}\tilde{\delta}_\sigma(r) \vec{S}_c \cdot \vec{S}_{\bar{c}} +\left[\frac{2 \alpha _s}{m_c^2r^3}-\frac{b}{2m_c^2r}\right]\vec{L} \cdot \vec{S}+\frac{ 4\alpha _s}{m_c^2r^3}\vec{T},
\end{equation}
\begin{equation}
V_{MNR}= -\frac{4}{3}\frac{\alpha _s}{r}+b\,r+\frac{32\pi \alpha _s}{9m_c^2} \tilde{\delta}_\sigma(r)\vec{S} _c \cdot \vec{S} _{\bar{c}} +\left[\frac{2 \alpha _s}{m_c^2r^3}+\frac{(4\varepsilon-1)b}{2m_c^2r}\right]\vec{L} \cdot \vec{S}+\left[\frac{ \alpha _s}{3m_c^2r^3}+\frac{\varepsilon b}{12m_c^2r}\right] \vec{T},
\end{equation}
where $\vec{L}$ is the orbital momentum and $\vec{S}$ is the spin of charmonium. In the mixed-confining model, $\varepsilon$ stands for the vector exchange scale. The singularity of contact hyperfine interaction within the spin-spin term has been smeared by Gaussian as in Ref.\cite{prd72}, $\tilde{\delta}_\sigma(r)=\left(\sigma/\sqrt{\pi}\right)^3e^{-\sigma^2r^2}$. The involved operators are diagonal in a $\mid\vec{J},\vec{L},\vec{S}\rangle$ basis with the matrix elements,
\begin{equation}
\langle \vec{S}_c\cdot\vec{S}_{\bar{c}} \rangle = \frac{1}{2} S(S+1)-\frac{3}{4},
\end{equation}
 \begin{equation}
 \langle \vec{L}\cdot\vec{S}\rangle = \frac{1}{2}\left[J(J+1)-L(L+1)-S(S+1)\right],
\end{equation}
\begin{equation}
\left\langle \vec{T}  \right\rangle = \left\langle \left[\frac{3}{r^2}(\vec{S}_c \cdot \vec{r})(\vec{S}_{\bar{c}} \cdot\vec{r})-(\vec{S}_c \cdot \vec{S}_{\bar{c}} )\right] \right\rangle = -\frac{6 \left(\langle \vec{L}\cdot\vec{S}\rangle\right)^2 + 3 \langle \vec{L}\cdot\vec{S}\rangle - 2S(S+1)L(L+1)}{6(2L-1)(2L+3)}.
\end{equation}
Instead of separating the spin-dependent interactions into leading order portions, we solved the Schroedinger equation of the unperturbed Hamiltonian with full potentials,
\begin{equation}
\left[-\frac{\hbar^2}{2\mu_R}\nabla^2+V_{c\bar{c}}(r)-E\right]\psi(\textbf{r})=0,
\end{equation}
where $\mu_R= m_c/2$ is the reduced mass. Here, $V_{c\bar{c}}(r)$ is $V_{NR}$ or $V_{MNR}$ which includes the spin-independent interactions as well as the spin-dependent ones. The Hamiltonian includes the full potential, enable us to maintain the subtleness of wave function. With the help of a well-chosen set of Gaussian basis functions, namely Gaussian Expansion Method \cite{gem}, the singular behavior of $1/r^3$ in spin-dependent terms at short distance can be refined variationally. Then, the wave functions $\psi_{lm}(\textbf{r})$ are expanded in terms of a set of Gaussian
basis functions as
\begin{equation}\label{wf1}
\psi_{lm}(\vec{r})=\sum^{n_{max}}_{n=1}C_{nl}\left(\frac{2^{2l+\frac{7}{2}}\nu_{n}^{l+\frac{3}{2}}}{\sqrt{\pi}(2l+1)!!}\right)^{\frac{1}{2}}
r^{l}e^{-(r/r_{n})^2}Y_{lm}(\hat{\vec{r}}),
\end{equation}
\begin{equation}\label{wf2}
\nu_{n}=\frac{1}{r_n^2}, \hspace{0.50cm}r_n=r_1a^{n-1}.
\end{equation}

The basis-related parameters $\{n_{max},r_1,r_n\}$ are determined by the variational principle, resulting the reasonably stable eigen solutions and differ slightly according to the explicit forms of potential models. Table \ref{basis} shows the utilized basis for the two potential models.

\begin{table}
\caption{Basis-related parameters for the potential models}
\centering
\label{basis}
\begin{tabular}{llcc}
\hline\noalign{\smallskip}
 &                           Parameter  &       NR       &   MNR    \\[3pt]
 \tableheadseprule\noalign{\smallskip}
\multirow{3}*{Basis Space}& $n_{max}$                 &      10    &     8    \\
&$r_1$ [GeV$^{-1}$]        &      0.4   &   0.4    \\
&$r_{n_{max}}$ [GeV$^{-1}$]&      15.8  &   6.8    \\
\noalign{\smallskip}\hline
\multirow{5}*{Potential Model}
&$m_c$ [GeV]               &1.4786      &     1.5216      \\
&$  \alpha_s $             & 0.5761     &     0.6344      \\
&$ b $ [GeV$^{2}$]         & 0.1468     &     0.1361      \\
&$\sigma  $ [GeV]          & 1.1384     &     1.2058      \\
&$\epsilon   $             &-           & -0.2193         \\
\noalign{\smallskip}\hline
\end{tabular}
\end{table}

To determine the parameters ($\vec{a}= a_1,a_2,\cdots,a_P$) appearing in the potentials, a merit function $\chi^2$ has been defined to search the best-fit parameters by its minimization,
\begin{equation}
\chi^2= \sum^{N}_{i=1}\left[\frac{M^{exp}_{i} - M^{th}_{i}(\vec{a})}{\sigma_i}\right]^2,
\end{equation}
where $N$ denotes the number of targeted data, the $\vec{\sigma}$ are the associated errors and $M_i$ represents the experimental and theoretical values. Given a trial set of model-depended parameters, a procedure calculating the $\chi^2(\vec{a})$ are developed to improve the trial solution with the increments $\delta \vec{a}$ and repeated until $\chi^2(\vec{a}+\delta \vec{a})$ effectively stops decreasing. The increments $\delta \vec{a}$ are solved by the set of linear equations
\begin{equation}
\sum^{P}_{j=1}\alpha_{kj}\delta a_j=\beta_k,
\end{equation}
with
\begin{eqnarray}
&& \beta_k\equiv -\frac{1}{2}\frac{\partial \chi^2}{\partial a_k} \\
&& \alpha_{kj}=\sum^{N}_{i=1}\frac{1}{\sigma^2_i}\left[\frac{\partial M^{th}_{i}(\vec{a})}{\partial a_k}\frac{\partial M^{th}_{i}(\vec{a})}{\partial a_j}\right].
\end{eqnarray}
The iteration has been taken in the numerical analysis of nonlinear systems. Here, the partial derivative has to be solved numerically because the $M^{th}_{i}(\vec{a})$ is not analytic.

\section{Numerical Results}
\label{sec:3}

\subsection{Mass Spectrum}
\label{sec:3.1}
Combining the nonrelativistic kinetic term and the interactions terms, we diagonalize the full Hamiltonian, which leads to a generalized eigenvalue problem. After the fitting procedure, the energy levels and corresponding eign functions of the two potential models are obtained with its optimized parameter set (shown in Table \ref{basis}).

The vector component of linear confining are fitted to be $21.93\%$ for the MNR model. This result consists with the one-fifth vector exchange given by Ref. \cite{prd75}. If defined an additional $r$-independent adjustable parameter in the vector linear confinement, the value of mixing coefficient should be $-1$ for some decay considerations \cite{DEbert03}. In addition, the scalar-vector mixing model requires less linear potential slope, but more splitting scale from spin-spin interaction.

Totally 60 states have been calculated with the two potential models, and summarized in Table \ref{cc}. We fit the mass of well-established states marked by $\ast$. Except somewhat low at $\chi _0(1^3P_0)$, both of the two potential models are overall good to reproduce the spectrum.

\begin{table}
\caption{ Experimental and theoretical Charmonium mass spectrum. The masses are in units of MeV, and the $\ast$ denotes the states used in the optimization of potential parameters. Our full-potential calculation results are listed in comparison with the perturbative results of NR model and its relativized extension GI model \cite{prd72}. }
\centering
\label{cc}
\begin{tabular}{lcccccc}
\hline\noalign{\smallskip}
          State                & Expt. \cite{pdg2010}     & \multicolumn{2}{c}{Ref. \cite{prd72}}     &        \multicolumn{2}{c}{Ours}  &\\
                               &                          &   NR       &  GI  &         NR       &    MNR     &\\[3pt]
\tableheadseprule\noalign{\smallskip}
   $\eta _c(1^1 S_0)^\ast$     &   $2980.3\pm1.2$         &  2982      & 2975 &        2990.4    &    2978.4  &\\
   $J/\Psi(1^3S_1)^\ast$       & $3096.916\pm0.011$       &  3090      & 3098 &        3085.1    &    3087.7  &\\
   $\eta^{\prime}_c(2^1 S_0)^\ast$& $3637\pm4$            &   3630     & 3623 &        3646.5    &    3646.9  &\\
   $\psi^{\prime}(2^3S_1)^\ast$& $3686.09\pm0.04$         &  3672      & 3676 &        3682.1    &    3684.7  &\\
   $\eta^{\prime}_c(3^1 S_0)$  &                          &   4043     & 4064 &        4071.9    &    4058.0  &\\
   $\psi^{\prime}(3^3S_1)^\ast$&     $4039\pm1$           &  4072      & 4100 &        4100.2    &    4087.0  &\\
   $\eta^{\prime}_c(4^1 S_0)$  &                          & 4384       & 4425 &        4420.9    &    4391.4  &\\
   $\psi^{\prime}(4^3S_1)^\ast$&    $4421\pm4$            &  4406      & 4450 &        4439.4    &    4411.4  &\\[3pt]
   $\chi _2 (1^3 P_2)^\ast$    &  $3556.20\pm0.09$        &  3556      & 3550 &        3551.4    &    3559.3  &\\
   $\chi _1(1^3P_1)^\ast$      &   $3510.66\pm0.07$       &  3505      & 3510 &        3500.4    &    3517.7  &\\
   $\chi _0(1^3P_0)^\ast$      &  $3414.75\pm0.31$        &  3424      & 3445 &        3351.9    &    3366.3  &\\
   $h_c (1^1P_1)^\ast$         &   $3525.42\pm0.29$       &  3516      & 3517 &        3514.6    &    3526.9  &\\
   $\chi _2 (2^3 P_2)$         &                          &  3972      & 3979 &        3979.8    &    3973.1  &\\
   $\chi _1(2^3P_1)$           &                          &  3925      & 3953 &        3933.5    &    3935.0  &\\
   $\chi _0(2^3P_0)$           &                          &  3852      & 3916 &        3835.7    &    3842.7  &\\
   $h_c (2^1P_1)$              &                          &  3934      & 3956 &        3944.6    &    3941.9  &\\
   $\chi _2 (3^3 P_2)$         &                          &  4317      & 4337 &        4383.4    &    4352.4  &\\
   $\chi _1(3^3P_1)$           &                          &  4271      & 4371 &        4317.9    &    4298.7  &\\
   $\chi _0(3^3P_0)$           &                          &  4202      & 4292 &        4216.7    &    4207.6  &\\
   $h_c (3^1P_1)$              &                          &  4279      & 4318 &        4333.9    &    4309.7  &\\
   $\chi _2 (4^3 P_2)$         &                          &            &      &        4736.7    &    4703.1  &\\
   $\chi _1(4^3P_1)$           &                          &            &      &        4620.3    &    4590.5  &\\
   $\chi _0(4^3P_0)$           &                          &            &      &        4551.8    &    4521.7  &\\
   $h_c (4^1P_1)$              &                          &            &      &        4639.5    &    4606.7  &\\[3pt]
   $\psi_3(1^3D_3)$            &                          &  3806      & 3849 &        3814.6    &    3812.6  &\\
   $\psi_2(1^3D_2)$            &                          &  3800      & 3838 &        3807.7    &    3820.1  &\\
   $\psi(1^3D_1)^\ast$         &     $3772.92\pm0.35$     &  3785      & 3819 &        3785.3    &    3808.8  &\\
   $\eta_{c2}(1^1D_2)$         &                          &  3799      & 3837 &        3807.3    &    3815.1  &\\
   $\psi_3(2^3D_3)$            &                          &  4167      & 4217 &        4182.9    &    4166.1  &\\
   $\psi_2(2^3D_2)$            &                          &  4158      & 4208 &        4173.7    &    4168.7  &\\
   $\psi(2^3D_1)^\ast$         &     $4153\pm3$           &  4142      & 4194 &        4150.4    &    4154.4  &\\
   $\eta_{c2}(2^1D_2)$         &                          &  4158      & 4208 &        4173.7    &    4164.9  &\\
   $\psi_3(3^3D_3)$            &                          &            &      &        4572.5    &    4526.5  &\\
   $\psi_2(3^3D_2)$            &                          &            &      &        4558.8    &    4523.6  &\\
   $\psi(3^3D_1)$              &                          &            &      &        4525.8    &    4502.2  &\\
   $\eta_{c3}(3^1D_2)$         &                          &            &      &        4559.7    &    4521.4  &\\[3pt]
   $\chi _4 (1^3 F_4)$         &                          &    4021    & 4095 &        4037.4    &    4024.7  &\\
   $\chi _3(1^3F_3)$           &                          &    4029    & 4097 &        4044.0    &    4047.6  &\\
   $\chi _2(1^3F_2)$           &                          &    4029    & 4092 &        4042.4    &    4059.7  &\\
   $h_{c3} (1^1F_3)$           &                          &    4026    & 4094 &        4041.1    &    4040.8  &\\
   $\chi _4 (2^3 F_4)$         &                          &    4348    & 4425 &        4371.1    &    4344.7  &\\
   $\chi _3(2^3F_3)$           &                          &    4352    & 4426 &        4374.4    &    4362.4  &\\
   $\chi _2(2^3F_2)$           &                          &    4351    & 4422 &        4369.9    &    4369.8  &\\
   $h_{c3} (2^1F_3)$           &                          &    4350    & 4424 &        4372.3    &    4356.8  &\\
   $\chi _4 (3^3 F_4)$         &                          &            &      &        4744.4    &    4684.8  &\\
   $\chi _3(3^3F_3)$           &                          &            &      &        4747.7    &    4698.5  &\\
   $\chi _2(3^3F_2)$           &                          &            &      &        4743.9    &    4704.2  &\\
   $h_{c3} (3^1F_3)$           &                          &            &      &        4745.9    &    4694.3  &\\[3pt]
   $\psi_5(1^3G_5)$            &                          &  4214      & 4312 &        4236.9    &    4213.5  &\\
   $\psi_4(1^3G_4)$            &                          &  4228      & 4320 &        4250.6    &    4244.7  &\\
   $\psi_3(1^3G_3)$            &                          &  4237      & 4323 &        4258.2    &    4267.4  &\\
   $\eta_{c4}(1^1G_4)$         &                          &  4225      & 4317 &        4247.1    &    4237.9  &\\
\noalign{\smallskip}\hline
\end{tabular}
\end{table}

\subsection{Leptonic Decays}
\label{sec:3.2}
The lowest-order expressions of electronic decay width the first-order QCD corrections \cite{prd37} are
\begin{equation}\label{eeS}
\Gamma_{ee} (nS) = \frac{4 \alpha^2 e^2_c}{M^2_{nS}}\mid R_{nS}(0)\mid^2\left(1-\frac{16}{3}\frac{\alpha_s(m_c)}{\pi}\right),
\end{equation}

\begin{equation}\label{eeD}
\Gamma_{ee} (nD) = \frac{25 \alpha^2 e^2_c}{2 M^2_{nD} M^4_{Q}}\mid R^{\prime\prime}_{nD}(0)\mid^2\left(1-\frac{16}{3}\frac{\alpha_s(m_c)}{\pi}\right),
\end{equation}
where $e_c$ is the c-quark charge in units of $\mid e\mid$, $\alpha = 1/137.036$ is the fine-structure constant, $M_{nS}$ and $M_{nD}$ are the $(n_r+1)th$ S-wave and D-wave state mass respectively with the radial excitation number $n_r$. Note that here $\alpha_s(m_c)$ and the $\alpha_s$ are essentially the strong coupling constants of different mass scales, and we adopt $\alpha_s(m_c) = 0.26$ as in Ref. \cite{screen03,DEbert2003twophoton}. $R_{nS}(0)$ is the radial $S$ wave function at the origin, and $R^{\prime\prime}_{nD}(0)$ is the second derivative of the radial $D$-wave function at the origin. Within the Gaussian basis space, the analytic formula for $R(0)$ is explicitly presented in the APPENDIX.
Table \ref{ee} compares the leptonic decay given in Ref. \cite{screen03} with our results in NR and MNR models.

\begin{table}
\caption{Leptonic decay width, in units of keV.}
\centering
\label{ee}
\begin{tabular}{cccccccccc}
\hline\noalign{\smallskip}
Particle &State & \multicolumn{2}{c}{Ref.\cite{screen03}}& Ref.\cite{SD2004}&  \multicolumn{2}{c}{NR}      &   \multicolumn{2}{c}{MNR}          &    Expt.\cite{pdg2010} \\
&&$\Gamma^{0}_{ee}$&$\Gamma_{ee}$&$\Gamma^{0}_{ee}$&$\Gamma^{0}_{ee}$&$\Gamma_{ee}$&$\Gamma^{0}_{ee}$&$\Gamma_{ee}$&\\[3pt]
\tableheadseprule\noalign{\smallskip}
$J/\Psi$         &$1^3S_1$      & 11.8    &  6.6   &12.13  & 5.6    &  3.1                    &  6.0  & 3.3     &    5.6$\pm$0.14$\pm$0.02\\
$\psi^{\prime}$  &$2^3S_1$      & 4.29    &  2.4   &5.03   & 2.3    &  1.3                    &  2.2  &  1.2    &    2.4$\pm$0.07\\
$\psi(4040)$  &$3^3S_1$      & 2.53    &  1.42   &3.48 & 1.9    &  1.0                    &  1.8  &  0.98   &    0.86$\pm$0.07 \\
$\psi(4415)$  &$4^3S_1$      & 1.25    &  0.7  &2.63   & 1.3    &  0.70                   &  1.3  &  0.70   &    0.58$\pm$0.07\\
$\psi(3770)$           &$1^3D_1$      & 0.055   &  0.031   &0.056& 0.089  &  0.050                  &  0.079&  0.044  &    0.27$\pm$0.018\\
$\psi(4160)$           &$2^3D_1$      & 0.066   &  0.037  &0.096 & 0.15   &  0.084                  &  0.13 & 0.073   &    0.83$\pm$0.07\\
\noalign{\smallskip}\hline
\end{tabular}
\end{table}

\subsection{Two-photon Decay}
\label{sec:3.3}
The two-photon decay widths are important to identify the potential charmonium states, i.e. $X(3915)$, $Z(3930)$ etc. and the latest $X(4160)$, $X(4350)$. With the first-order QCD radiative corrections \cite{prd37}, the two-photon decay widths of $^1S_0$, $^3P_0$ and $^3P_2$ explicitly are

\begin{equation}
\Gamma_{\gamma \gamma}(^1S_0)=\frac{3 \alpha^2 e^4_c \mid R_{nS}(0)\mid^2}{m^2_c}\left[1+\frac{\alpha_s(m_c)}{\pi}\left(\frac{\pi^2}{3}
-\frac{20}{3}\right)\right],
\end{equation}
\begin{equation}
\Gamma_{\gamma \gamma}(^3P_0)=
\frac{27 \alpha^2 e^4_c \mid R^{\prime}_{nP}(0)\mid ^2}{m^4_c}\left[1+\frac{\alpha_s(m_c)}{\pi}\left(\frac{\pi^2}{3}-\frac{28}{9}\right)\right],
\end{equation}
\begin{equation}
\Gamma_{\gamma \gamma}(^3P_2)=\frac{36 \alpha^2 e^4_c \mid R^{\prime}_{nP}(0)\mid^2}{5 m^4_c}\left[1-\frac{16}{3}\frac{\alpha_s(m_c)}{\pi}\right],
\end{equation}
where $R^{\prime}_{nP}(0)$ is the first derivative of the radial $P$-wave function at the origin. Our numerical results of two-photon decay widths are shown in Table \ref{gamma} in comparison with Refs.\cite{DEbert2003twophoton,Gupta1996}.

\begin{table}
\caption{ Two-photon decay width, in units of keV.}
\centering
\label{gamma}
\begin{tabular}{lccccc}
\hline\noalign{\smallskip}
State        & Ref. \cite{DEbert2003twophoton}& Ref. \cite{Gupta1996}&  NR      &  MNR     &    Expt. \cite{pdg2010} \\[3pt]
\tableheadseprule\noalign{\smallskip}
$1^1S_0$     &   5.5    &    10.94   &   7.4              &  7.5   &    6.7$^{+0.9}_{-0.8}$\\
$2^1S_0$     &   1.8    &            &   3.2              &  2.9   &    \\
$3^1S_0$     &          &            &   2.9              &  2.5   &      \\
$4^1S_0$     &          &            &   2.0              &  1.8   &     \\
$1^3P_0$     &  2.9     &   6.38     &   11               &  10.8  &    2.29$\pm$0.18\\
$2^3P_0$     &  1.9     &            & 7.7                & 6.7    &     \\
$3^3P_0$     &          &            &  7.9               & 6.5    &    \\
$1^3P_2$     &  0.50    &    0.57    &   0.29             &  0.27  &    0.50$\pm$0.03\\
$2^3P_2$     &  0.52    &            &   0.43             & 0.39   &     \\
$3^3P_2$     &          &            &   0.81             &  0.66  &    \\
\noalign{\smallskip}\hline
\end{tabular}
\end{table}

\subsection{Radiative Transition}
\label{sec:3.4}
Because radiative transition is sensitively dependent on the detailed features of the wave functions, it is of great interest to have careful inspection. Besides, it has been noticed that the known M1 rates showing serious disagreement between the previous theoretical calculation \cite{prd72} and experiment. In our full-potential calculation framework, the wave functions are directly associated with the eigenvectors of Hamiltonian corresponding to the masses. The E1 transition rate between an initial charmonium state $i$ of radial quantum number $n_i$, orbital angular momentum $L_i$, spin $S_i$, and total angular momentum $J_i$, and a final state $f$ is given by Ref. \cite{Kwong1988} as
 \begin{equation}\label{e1}
\Gamma_{E1}\left(n_i^{2S_i+1}L_{i_{J_i}}\rightarrow  n_f^{2S_f+1}L_{f_{J_f}}\right)=\frac{4}{3}C_{fi}\,\delta_{S_iS_f}e^2_c\,\alpha  \left|\left\langle \psi_f \left|r\right| \psi_i \right\rangle \right|^2
E^3_{\gamma}\frac{E_f^{(c\bar{c})}}{M_i^{(c\bar{c})}} ,
\end{equation}
\begin{equation}\label{m1}
\Gamma_{M1}\left(n_i^{2S_i+1}L_{i_{J_i}}\rightarrow   n_f^{2S_f+1}L_{f_{J_f}}\right) =\frac{4}{3}\frac{2J_f+1}{2L_i+1}e^2_c\,\frac{\alpha}{m_c^2}\delta_{L_iL_f} \delta_{S_i,S_f\pm 1}  \left|\left\langle \psi_f \mid \psi_i \right\rangle \right|^2
E^3_{\gamma}\frac{E_f^{(c\bar{c})}}{M_i^{(c\bar{c})}}.
\end{equation}
In the above formulas, $e_c$ is the charge of c-quark in unit of $\mid e\mid$, and $M_i$,$E_f$ represent the eigen mass of initial state and the total energy of final state respectively. The momentum of the final photon equals $E_{\gamma}=(M^2_i-M^2_f)/(2M_i)$ in the nonrelativistic approximation \cite{QWG}. The angular matrix element  $C_{fi}$ is
\begin{equation}
C_{fi}=max(L_i,L_f)(2J_f+1)\left\{
                 \begin{array}{clr}
                  L_f & J_f & S \\
                  J_i  & L_i  & 1
                 \end{array} \right\}^2.
\end{equation}
The Gaussian expanded wave functions give rise to the analytic formulas for the overlap integral, i.e. Eq. (\ref{em1}), and the transition matrix elements Eq. (\ref{em2}). The $n_{max}$ stands for the dimension of Gaussian basis as defined above. The numerical results of E1 transition rates are presented in Table \ref{ccE1-1} and Table \ref{ccE1-2}, as well as those of M1 transition in Table \ref{ccM1}.

\begin{table}
\caption{2S, 3S, 1P and 2P E1 radiative transitions}
\centering
\label{ccE1-1}
\begin{tabular}{lcccccccccc}
\hline\noalign{\smallskip}
               &              &          \multicolumn{4}{c}{Ref.\cite{prd72}}    & \multicolumn{4}{c}{Ours}  & $\Gamma _{expt.}$\cite{pdg2010} [keV]\\
 Initial meson & Final meson  &   \multicolumn{2}{c}{$E_{\gamma}$ [MeV]}   &   \multicolumn{2}{c}{$\Gamma_{thy}$ [keV]}      & \multicolumn{2}{c}{$E_{\gamma}$ [MeV]}           &  \multicolumn{2}{c}{$\Gamma_{thy}$ [keV]}  &  \\

                         &                        &                           NR & GI &NR &GI   &NR  &MNR&NR  &MNR &   \\[3pt]
\tableheadseprule\noalign{\smallskip}
                            $\psi^{\prime}(2^3S_1)$&$\chi_2(1^3P_2)$         & 128&128 &38 &24   &128 &123&43  &39  &$26.6\pm1.1$\\
                                                   &$\chi_1(1^3P_1)$         & 171&171 &54 &29   &177 &163&48  &38  &$28.0\pm1.2$\\
                                                   &$\chi_0(1^3P_0)$         & 261&261 &63 &26   &315 &305&34  &29  &$29.2\pm0.9$\\
                           $\eta^{\prime}_c(2^1S_0)$ &$h_c(1^1P_1)$          & 111&119 &49 &36   &130 &118&72  &56 &\\
                            $\psi^{\prime}(3^3S_1)$&$\chi_2(1^3P_2)$         & 455&508 &0.70&12.7&512 &494&6   &6.44  &$<1.4$\\
                                                   &$\chi_1(1^3P_1)$         & 494&547 &0.53&0.85&556 &530&0.76&0.44  &$<0.9$\\
                                                   &$\chi_0(1^3P_0)$         & 577&628 &0.27&0.63&680 &657&8.9 &8.1  &\\
                           $\eta^{\prime}_c(3^1S_0)$ &$h_c(1^1P_1)$          & 485&511 &9.1 &28  &519 &496&6.1 &7.7  &\\
                           $\chi _2(1^3P_2) $      &$J/\psi(1^3S_1)$         & 429&429 &424&313 &436 &440&421  &405 &$384.2\pm16$\\
                            $\chi _1(1^3P_1)$      &                         & 390&389 &314&239 &391 &404&330  &341 &$295.8\pm13$ \\
                            $\chi _0(1^3P_0)$      &                         & 303&303 &152&114 &256 &267&97   &104 &$119.5\pm8$ \\
                            $h_c (1^1P_1)$         &$\eta _c (1^1S_0)$       & 504&496 &498&352 &485 &506&465  &473 & \\
                            $\chi _2 (2^3 P_2)$    &$\psi ^{\prime}(2^3S_1)$ & 276&282 &304&207 &287 &278&300  &264 & \\
                            $\chi _1(2^3P_1)$      &                         & 232&258 &183&183 &243 &242&243  &234 & \\
                            $\chi _0(2^3P_0)$      &                         & 162&223 &64 &135 &151 &155&77   &83  & \\
                            $h_c(2^1P_1)$          &$\eta^{\prime}_c(2^1S_0)$& 285&305 &280&218 &287 &284&297  &274 & \\
                            $\chi_2(2^3P_2)$       &$J/\psi(1^3S_1)$         & 779&784 &81 &53  &794 &787&108  &111 & \\
                            $ \chi_1(2^3P_1)$      &                         & 741&763 &71 &14  &757 &756&27   &33  & \\
                            $\chi_0 (2^3P_0)$      &                         & 681&733 &56 &1.3 &677 &681&30   &28  & \\
                            $h_c(2^1P_1)$          &$\eta _c(1^1S_0)$        & 839&856 &140&85  &839 &846&104  &116 & \\
                            $\chi_2(2^3P_2)$       &$\psi_3(1^3D_3)$         & 163&128 &88 &29  &162 &157&81   &76  & \\
                                                   &$\psi_2(1^3D_2)$         & 168&139 &17 &5.6 &168 &150&14   &10  &  \\
                                                   & $\psi(1^3D_1)$          & 197&204 &1.9&1.0 &190 &161&0.98 &0.64& \\
                            $\chi_1(2^3P_1)$       & $\psi_2(1^3D_2)$        & 123&113 &35 &18  &124 &113&37   &30  & \\
                                                   & $\psi(1^3D_1)$          & 152&179 &22 &21  &145 &124&16   &11  & \\
                            $\chi_0(2^3P_0)$       & $\psi(1^3D_1)$          & 81 &143 &13 &51  &50  &34 &4.2  &1.4 & \\
                            $h_c(2^1P_1)$          & $\eta_{2c}(1^1D_2)$     & 133&117 &60 &27  &135 &125&61   &51  &  \\
   \noalign{\smallskip}\hline
\end{tabular}
\end{table}

\begin{table}
\caption{3P, 1D and 2D E1 radiative transitions }
\centering
\label{ccE1-2}
\begin{tabular}{lcccccccccc}
\hline\noalign{\smallskip}
               &          &          \multicolumn{4}{c}{Ref. \cite{prd72}}    & \multicolumn{4}{c}{Ours}  & $\Gamma _{expt.}$ \cite{pdg2010} \\
 Initial meson & Final meson  &   \multicolumn{2}{c}{$E_{\gamma}$ [MeV]}   &   \multicolumn{2}{c}{$\Gamma_{thy}$ [keV]}      & \multicolumn{2}{c}{$E_{\gamma}$ [MeV]}           &  \multicolumn{2}{c}{$\Gamma_{thy}$ [keV]}  & [keV] \\
                          &                        &                           NR & GI &NR &GI   &NR  &MNR&NR  &MNR &   \\[3pt]
\tableheadseprule\noalign{\smallskip}
                            $\chi_2(3^3P_2)$       &$\psi(3^3S_1)$           & 268&231 &509&199 &273 &257&434  &360       & \\
                            $ \chi_1(3^3P_1)$      &                         & 225&212 &303&181 &212 &206&292  &272      & \\
                            $\chi_0 (3^3P_0)$      &                         & 159&188 &109&145 &115 &119&63   &72      & \\
                            $h_c(3^1P_1)$          &$\eta _c(3^1S_0)$        & 229&246 &276&208 &254 &244&377  &332       & \\
                            $\chi _2 (3^3 P_2)$    &$\psi ^{\prime}(2^3S_1)$ & 585&602 & 55&30  &645 &617&112  &97       & \\
                            $\chi _1(3^3P_1)$      &                         & 545&585 & 45&8.9 &589 &570&30   &31       & \\
                            $\chi _0(3^3P_0)$      &                         & 484&563 &32 &0.045 &501&490&25  &19      & \\
                            $h_c(3^1P_1)$          &$\eta^{\prime}_c(2^1S_0)$& 593&627 & 75& 43 &633 &612&95   &90       & \\
                            $\chi _2(3^3P_2) $      &$J/\psi(1^3S_1)$         &1048&1063 & 34& 19&1106&1081&62  &59      & \\
                            $\chi _1(3^3P_1)$      &                         &1013&1048 & 31&2.2&1057&1040&10  &11       & \\
                            $\chi _0(3^3P_0)$      &                         & 960&1029 & 27&1.5&980 &971 &23  &22      & \\
                            $h_c (3^1P_1)$         &$\eta _c (1^1S_0)$       &1103&1131 & 72& 38&1135&1126&73  &76      & \\

                            $\chi_2(3^3P_2)$       &$\psi_3(2^3D_3)$         & 147&118  &148& 51&196 &182 &279 &231      & \\
                                                   &$\psi_2(2^3D_2)$         & 156&128  & 31&9.9&205 &180 &48  &34      & \\
                                                  & $\psi(2^3D_1)$          & 155&141  &2.1&0.77&227&194 &3.2 &2.1       & \\
                            $\chi_1(3^3P_1)$       & $\psi_2(2^3D_2)$        & 112&108  &58 &35 &142 &128 &121 &92       & \\
                                                   & $\psi(2^3D_1)$          & 111&121  &19 &15 &164 &142 &50  &34      & \\
                            $\chi_0(3^3P_0)$       & $\psi(2^3D_1)$          & 43 & 97  &4.4&35 &66  &53  &21  &11       & \\
                            $h_c(3^1P_1)$          & $\eta_{2c}(2^1D_2)$     & 119&109  &99 &48 &157 &142 &205 &158      & \\
                            $\chi_2(3^3P_2)$       &$\psi_3(1^3D_3)$         & 481&461  &0.049&6.8&532&506&1.6 &1.4       & \\
                                                   &$\psi_2(1^3D_2)$         & 486&470  &0.0091&0.13&538&500&0.76&0.42     & \\
                                                   & $\psi(1^3D_1)$          & 512&530  &0.00071&0.001&557&510&0.11&0.062     & \\
                            $\chi_1(3^3P_1)$       & $\psi_2(1^3D_2)$        & 445&452  &0.035&4.6 &480&452&1.2 &1.8       & \\
                                                   & $\psi(1^3D_1)$          & 472&512   &0.014&0.39&500&462&0.11 &0.00029    & \\
                            $\chi_0(3^3P_0)$       & $\psi(1^3D_1)$          & 410&490   &0.037&9.7&409&380&28  &30       & \\
                            $h_c(3^1P_1)$          & $\eta_{2c}(1^1D_2)$     & 453&454   &0.16&5.7&495 &466&0.21&0.56       & \\

                            $\psi_3(1^3D_3)$        &$\chi_2(1^3P_2)$         & 242&282 &272&296 &254 &245&340 &302 &\\
                            $\psi_2(1^3D_2)$        &$\chi_2(1^3P_2)$         & 236&272 &64 &66 &248 &252&79   &82  &\\
                                                   &$\chi_1(1^3P_1)$         & 278&314 &307&268 &295 &290&321 &301 &\\
                            $\psi(1^3D_1)$          &$\chi_2(1^3P_2)$         & 208&208 &4.9&3.3 &227 &241&6.8 &8.1  &$<24.6$\\
                                                   &$\chi_1(1^3P_1)$         & 250&251 &125&77  &274 &280&146 &153  &$79.2\pm16$\\
                                                   &$\chi_0(1^3P_0)$         & 338&338 &403&213 &409 &417&367 &362 &$199.3\pm25$\\
                            $h_c(1^1D_2)$          &$h_c(1^1P_1)$            & 264&307 &339&344 &281 &277&398 &374 &  \\
                            $\psi_3(2^3D_3)$        &$\chi_2(1^3P_2)$         & 566&609 & 29& 16 &584 &563&23  &23  &\\
                            $\psi_2(2^3D_2)$        &$\chi_2(1^3P_2)$         & 558&602 &7.1&0.62&576 &565&0.85&1.5 &\\
                                                   &$\chi_1(1^3P_1)$         & 597&640 & 26&23  &619 &600&38  &41  &\\
                            $\psi(2^3D_1)$          &$\chi_2(1^3P_2)$         & 559&590 &0.79&0.027&556&552&0.18&0.080&$<1.3$\\
                                                   &$\chi_1(1^3P_1)$         & 598&628 & 14&3.4 &599 &588&3.6 &5.5 &$<0.7$\\
                                                   &$\chi_0(1^3P_0)$         & 677&707 & 27& 35 &722 &713&99  &109 &\\
                            $h_c(2^1D_2)$          &$h_c(1^1P_1)$            & 585&634 & 40& 25 &607 &589&38  &41  &\\
\noalign{\smallskip}\hline
\end{tabular}
\end{table}


\begin{table}
\caption{M1 radiative partial widths }
\centering
\label{ccM1}
\begin{tabular}{lcccccccccc}
\hline\noalign{\smallskip}
 Initial & Final   & \multicolumn{4}{c}{Ref. \cite{prd72}}  &  \multicolumn{4}{c}{Ours} & $\Gamma _{expt.}$ [keV]\\
   meson & meson   &  \multicolumn{2}{c}{$E_{\gamma}$ [MeV]}   &  \multicolumn{2}{c}{$\Gamma_{thy}$ [keV]}  & \multicolumn{2}{c}{$E_{\gamma}$ [MeV]}             &  \multicolumn{2}{c}{$\Gamma_{thy}$ [keV]}     &\\
               &                           &                            NR  & GI & NR  & GI  &NR   &MNR& NR   & MNR &        \\[3pt]
\tableheadseprule\noalign{\smallskip}
            $J/\psi(1^3S_1)$               &$\eta _c (1^1S_0)$         & 116 &115 & 2.9 & 2.4 &93   &107&1.5   &2.2  &$1.58\pm0.37$ \cite{pdg2010}  \\
            $\psi ^{\prime}(2^3S_1)$       &$\eta^{\prime}_c(2^1S_0)$  & 48  & 48 &0.21 &0.17 &35   &38 &0.086 &0.096&$0.143\pm0.027\pm0.092$ \cite{BES2011} \\
                                           &$\eta _c (1^1S_0)$         & 639 &638 & 4.6 & 9.6 &627  &639&3.1   &3.8  &$0.97\pm0.14$ \cite{2Sto1S2009}\\
            $\eta^{\prime}_c(2^1S_0)$      &$J/\psi(1^3S_1)$           & 501 &501 & 7.9 & 5.6 &518  &516&6.1   &6.9  &          \\
            $\psi(3^3S_1)$                 &$\eta_c(3^1S_0)$           & 29  & 35 &0.046&0.067&28   &29 &0.043 &0.044&          \\
                                           &$\eta^{\prime}_c(2^1S_0)$  & 382 &436 & 0.61& 2.6 &429  &416&0.70  &0.71 &          \\
                                           &$\eta _c (1^1S_0)$         & 922 &967 & 3.5 & 9.0 &960  &958&3.2   &3.7  &          \\
            $\eta_c(3^1S_0)$               &$\psi^{\prime}(2^3S_1)$    & 312 &361 & 1.3 &0.84 &371  &356&1.7   &1.6  &          \\
                                           &$J/\psi(1^3S_1)$           & 810 &856 & 6.3 & 6.9 &867  &854&5.9   &6.5  &          \\
            $h^{\prime}_c(2^1P_1)$         &$\chi_2(1^3P_2)$           & 360 &380 &0.071&0.11 &374  &364&1.3   &1.2  &          \\
                                           &$\chi_1(1^3P_1)$           & 400 &420 &0.058&0.36 &419  &401&0.16  &0.13 &          \\
                                           &$\chi_0(1^3P_0)$           & 485 &504 &0.033& 1.5 &548  &534&5.6   &5.3  &          \\
             $\chi _2(2^3P_2)$             &$h_c(1^1P_1)$              & 430 &435 &0.067& 1.3 &438  &421&1.0   &0.89 &          \\
             $\chi _1(2^3P_1)$             &$h_c(1^1P_1)$              & 388 &412 &0.050&0.045&397  &387&0.15  &0.13 &          \\
             $\chi_0(2^3P_0)$              &$h_c(1^1P_1)$              & 321 &379 &0.029&0.50 &308  &303&4.8   &4.8  &          \\
           $h^{\prime}_c(3^1P_1)$          &$\chi_2(2^3P_2)$           &     &    &     &     &340  &323&1.7   &1.5  &          \\
                                           &$\chi_1(2^3P_1)$           &     &    &     &     &382  &358&0.17  &0.12 &          \\
                                           &$\chi_0(2^3P_0)$           &     &    &     &     &470  &442&3.5   &2.7  &          \\
                                           &$\chi_2(1^3P_2)$           &     &    &     &     &712  &685&2.3   &2.1  &          \\
                                           &$\chi_1(1^3P_1)$           &     &    &     &     &753  &719&0.27  &0.22 &          \\
                                           &$\chi_0(1^3P_0)$           &     &    &     &     &871  &840&8.9   &8.2  &          \\
             $\chi _2(3^3P_2)$             &$h_c(2^1P_1)$              &     &    &     &     &417  &391&1.2   &0.99 &          \\
                                           &$h_c(1^1P_1)$              &     &    &     &     &783  &747&1.8   &1.5  &          \\
             $\chi _1(3^3P_1)$             &$h_c(2^1P_1)$              &     &    &     &     &357  &342&0.16  &0.12 &          \\
                                           &$h_c(1^1P_1)$              &     &    &     &     &729  &703&0.22  &0.20&          \\
             $\chi_0(3^3P_0)$              &$h_c(2^1P_1)$              &     &    &     &     &263  &257&3.7   &3.3  &          \\
                                           &$h_c(1^1P_1)$              &     &    &     &     &644  &626&5.3   &5.1  &          \\
\noalign{\smallskip}\hline
\end{tabular}
\end{table}

\section{Discussion}
\label{sec:4}

\subsection{$\psi(3686)$, $\psi(3770)$, $\psi(4160)$}
\label{sec:4.1}
The world-average $\Gamma_{ee}(\psi(3770))$ is $0.265\pm 0.018$ keV \cite{pdg2010}. The significant leptonic width implies that there is a sizeable S-D mixing between the $2^3S_1$ and $1^3D_1$ state, since it is expected to be highly suppressed if $\psi(3770)$ is a pure D-wave state (shown in Table \ref{ee}). The mixing arises both from the usual relativistic correction terms and coupling to strong decay channels, and will affect the E1 transition rates of $\psi(3770)$ and $\psi(3686)$ \cite{QWG}. We investigate the mixing under the present calculation framework from two scenarios respectively: electronic annihilation and dipole transition. Assuming the $\psi(3686)$ and $\psi(3770)$ to be a mixture of a $1^3D_1$ and a $2^3S_1$ state as
 \begin{eqnarray}\label{mix1}
\mid\left.\psi(3686) \right\rangle &=& \mid \left.2^3S_1\right\rangle \cos\theta - \mid\left.1^3D_1\right\rangle \sin\theta,   \nonumber \\
\mid\left.\psi(3770) \right\rangle &=& \mid\left.2^3S_1\right\rangle \sin\theta + \mid\left.1^3D_1\right\rangle \cos\theta.
\end{eqnarray}
Then, Eqs. (\ref{eeS}, \ref{eeD}) can be expressed as

\begin{equation}\label{ee3686}
 \Gamma_{ee}(\psi(3686))= \frac{4 \alpha^2 e^2_c}{M^2_{\psi^{\prime}}}\left| R_{2S}(0)\cos \theta - \frac{5}{2\sqrt{2} m^2_c} R^{\prime\prime}_{1D}(0)\sin \theta \right|^2 ,
\end{equation}
\begin{equation}\label{ee3770}
\Gamma_{ee}(\psi(3770)) = \frac{4 \alpha^2 e^2_c}{M^2_{\psi^{\prime\prime}}}\left| R_{2S}(0) \sin \theta + \frac{5}{2\sqrt{2} m^2_c} R^{\prime\prime}_{1D}(0)\cos \theta \right|^2 .
\end{equation}

\begin{figure}
\centering
\includegraphics{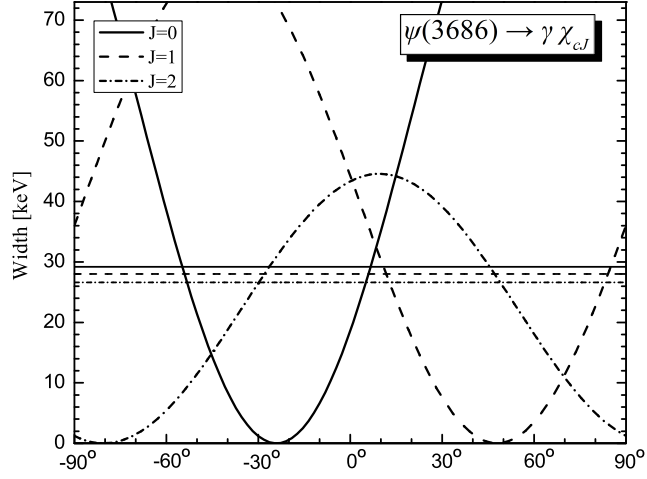}
\caption{The E1 radiative transition rate of $\psi(3686)\rightarrow \gamma 1^3P_J$ dependent on the various mixing angle $\theta$. Because of the relatively precise and very close measurements for the three channels, we illustrate the central values of $\Gamma_{expt.}$ \cite{pdg2010} with single lines rather than the error-associated areas.}
\label{mix3686}
\end{figure}

\begin{figure}
\centering
\includegraphics{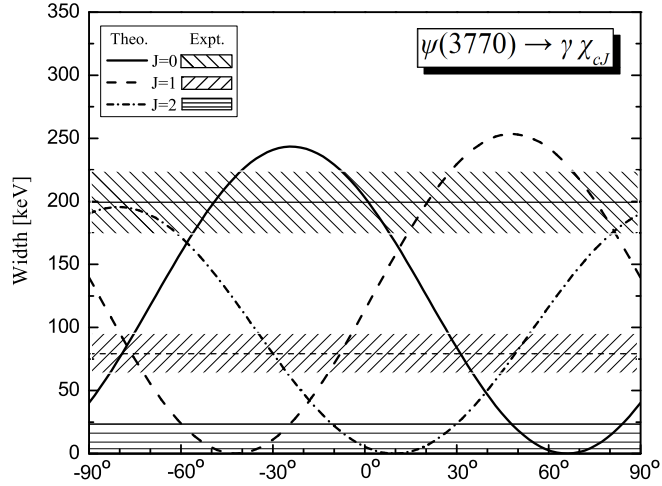}
\caption{The E1 radiative transition rate of $\psi(3770)\rightarrow \gamma 1^3P_J$ dependent on the various mixing angle $\theta$. The experimental data \cite{pdg2010} are illustrated by the central values as well as the related errors.}
\label{mix3770}
\end{figure}

Through the fitting of the ratio $\Gamma(\psi(3770) \rightarrow e^+e^-)/\Gamma(\psi(3686) \rightarrow e^+e^-)$, combining the experimental data \cite{pdg2010} and the radial wave function values at the origin, i.e. $R_{2S}(0)=0.556$ GeV$^{3/2}$ and $R^{\prime\prime}_{1D}(0)=0.144$ GeV$^{7/2}$,  we find two solutions for the mixing angle:
\begin{equation}
\theta =  8 ^\circ \pm 0.4^\circ   \; \;\;  \mathtt{or}   \;\;\; \theta =  -30 ^\circ   \pm 0.4 ^\circ.
\end{equation}
These solutions are smaller than $(12\pm 2)^\circ$ or $-(27\pm2)^\circ$ determined by leptonic decay widths in \cite{Rosner2001,screen03}, partially due to more precise measurements. Impending, we illustrate the mixing-dependency of E1 transitions of our model in Fig. \ref{mix3686} and Fig. \ref{mix3770}. The decay widths of $\psi(3686)\rightarrow \gamma \chi_{c0}$, $\gamma \chi_{c1}$ obviously favor $\theta =  8 ^\circ$, while the matching angle for $\gamma \chi_{c2}$ locates around $\theta =  -30 ^\circ$ and $\theta =  50 ^\circ$. The $\theta =  50 ^\circ$ would be too large for the two states mixing. On the other hand, a positive and slight mixing is compatible for $\psi(3770)\rightarrow \gamma \chi_{c0}$, $\gamma \chi_{c2}$ consisting with $\theta =  8 ^\circ$, except the challenging $\gamma \chi_{c1}$ fitting with a small but negative angle. This discrepancy reveals that $S$-$D$ wave mixing would be important for $\psi(3770)$ and $\psi(3686)$, but not sufficient. The mixing-angle functions for the two out-standing decays may be modified by the first-order relativistic corrections to their wave functions, since the relativistic corrections play an essential role in E1 radiative transitions indeed, as suggested in \cite{Rosner2001}.

The $S$-$D$ mixing may be more serious for the higher charmonium above the open-charm threshold \cite{4160chao,SD2004,screen03,SDmix01}. As shown in Table \ref{ee}, the experimental leptonic decay width of the $\psi(4160)$ is significantly larger than the potential model predictions as $2^3D_1$.  In Ref. \cite{SDmix01}, it is explored that the possible mixing in $(n+1)^3S_1$-$n ^3D_1$ determined by the ratio of di-electron widths and get a large mixing angel $\sim 34 ^\circ$ ($n=2,3$), while the hyperfine effects are neglected in the wave functions by a universal potential. In the current framework, the spin-dependent interactions at $\mathcal {O}(v^2/c^2)$ are thoroughly considered to determination of the state wave functions. Besides the distinct leptonic decay widths, it is found that the theoretical E1 radiative rates of the pure $2^3D_1$ and $3^3S_1$ are incomparable with the available experimental data. This discrepancy seems cannot be eliminated by the relativistic corrections, hence dissimilar with the situations of $\psi(3770)$ and $\psi(3686)$. We extend above pattern to the $\psi(4040)$-$\psi(4160)$ and $\psi(4160)$-$\psi(4415)$ mixing scenarios, and also independently via the leptonic decays as well as their E1 radiative rates. Combining the corresponding values of wave functions at the origin, i.e. $R_{3S}(0)=0.556$ GeV$^{3/2}$, $R_{4S}(0)=0.508$ GeV$^{3/2}$ and $R^{\prime\prime}_{2D}(0)=0.201$ GeV$^{7/2}$ and the central values of experimental data $\Gamma_{ee}(\psi(4040))=0.86\pm0.07$ keV, $\Gamma_{ee}(\psi(4160))=0.83\pm0.07$ keV and $\Gamma_{ee}(\psi(4415))=0.58\pm0.07$ keV \cite{pdg2010}, the mixing angles determined by the leptonic decay widths are listed in Table \ref{mix2}.

\begin{table}
\caption{The $3^3S_1$-$2 ^3D_1$ and $4^3S_1$-$2^3D_1$ mixing for the higher charmonium di-electron decays. Our predictions are obtained with the MNR potential model. Since the inverse definitions about mixing angle, these assignments in Ref. \cite{4160chao} are reasonably exchanged here for a proper comparison. }
\centering
\label{mix2}
\begin{tabular}{lcccc}
\hline\noalign{\smallskip}
 Mixing states                     & \multicolumn{3}{c}{Mixing angle}       \\
                                   & Ref.\cite{4160chao}& Ref.\cite{SDmix01}    & Ours &   \\[3pt]
\tableheadseprule\noalign{\smallskip}
$3S$-$2D$                 & $35^\circ/-55^\circ$ & $34.8^\circ/-55.7^\circ$ &  $29.8^\circ/-60.7^\circ$   \\
$4S$-$2D$                 &                      &                          &  $24.8^\circ/-58.5^\circ$   \\
\noalign{\smallskip}\hline
\end{tabular}
\end{table}

\begin{figure}
\centering
\includegraphics{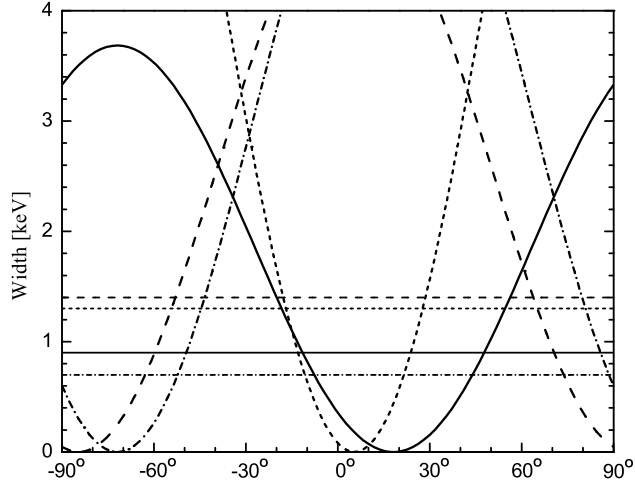}
\caption{The mixing angle $\theta$-dependent E1 radiative transition rates of $\psi(4040)\rightarrow \gamma \chi_{c1}$ (solid), $\psi(4040)\rightarrow \gamma \chi_{c2}$ (dash), $\psi(4160)\rightarrow \gamma \chi_{c1}$ (short dash dot) and $\psi(4160)\rightarrow \gamma \chi_{c2}$ (short dash). The experimental upper limits \cite{pdg2010} are represented in the same line-shape with the corresponding channels.}
\label{higherSD}
\end{figure}

In Fig.\ref{higherSD}, one can figure out that the mixing angles given by the $\Gamma_{ee}$, i.e. $29.8^\circ$ or $-60.7^\circ$, are still acceptable for the E1 transitions. Then, it is in a dilemma that the $\Gamma(\psi(4040)\rightarrow \gamma \chi_{c1})$ and $\Gamma(\psi(4160)\rightarrow \gamma \chi_{c2})$ favor $\theta = 29.8^\circ$, while $\Gamma(\psi(4040)\rightarrow \gamma \chi_{c2})$ and $\Gamma(\psi(4160)\rightarrow \gamma \chi_{c1})$ support $\theta = -60.7^\circ$. Both of the two solutions seems to be equally important, and it is difficult to rule out any one of them from the current perspective, possibly due to the mixing model being oversimplified. Nevertheless, it would be still interesting to glean something for the further understandings and revisiting with more precise measurements.

\subsection{$X(3872)$ and $X(3915)$} 
\label{sec:4.2}
The first surprising charmonium-like state $X(3872)$ \cite{3872expt} is still far away from sufficiently understanding. Its quantum numbers have not been decided yet from $J^{PC}=1^{++}$ and $J^{PC}=2^{-+}$ \cite{3872jpc}. The radiative decay $\gamma J/\psi$ \cite{3872gamma,3872belle1++} bolsters the $1^{++}$ assignment, while the $J^P=2^-$ is ratherish favored by a study of the $3\pi$ invariant mass distribution in $J/\psi \omega$ decays \cite{3872belle2-+,babar2010}. Assuming the charmonium structure, these $J^{PC}$ imply the possible candidates, i.e. $\chi_{c1}(2^3P_1)$ and $\eta_{c2}(1^1D_2)$. As the spectrum listed in Table \ref{cc}, the masses of the two states remain a gap around 50 MeV with the world-average mass $M=3871.56\pm0.22$ MeV \cite{pdg2010}. Combining the observed branching fraction \cite{3872gamma} and the full width, it is evaluated that the radiative decay widths are $\Gamma(X(3872)\rightarrow J/\psi \gamma) > 20.7$ keV and $\Gamma(X(3872)\rightarrow \psi(2S) \gamma) > 0.069$ MeV. The theoretical results of the radiative decays for the conventional charmonium candidates are apparently too large, and the radio of the two rates is nearly two times larger than the observation. Hence, the $c\bar{c}$ nature of $X(3872)$ seems unlikely in the present framework.

Another interesting resonance named $X(3915)$ has been announced by Belle Collaboration \cite{3915expt2010}, which is found in the processes $\gamma\gamma\rightarrow J/\psi \omega$. $X(3915)$ clearly has the positive $C$-parity, but the $J^P$ assignment remains to be determined. Owning to the same production and decay signatures, the $Y(3940)$ \cite{3940Ybelle,3490Ybabar2008} and the $X(3915)$ are considered to be the same state prevailingly,
which is close to those of the $Z(3930)$. If the $Z(3930)$ (discussed in the coming section) occupies the assignment of $\chi^{\prime}_{c2}$ \cite{prd32,prd69}, the possibility of $X(3915)$ can be ruled out, or else they are the same state as $\chi_{c2}(2^3P_2)$. The world-average mass and width of the $X(3915)$ are $M=3917.4\pm2.7$ MeV, $\Gamma=28^{+10}_{-9}$ MeV \cite{pdg2010}. In our predictions, the mass of $\chi_{c2}(2^3P_2)$ is about 60 MeV higher than the experimental data for the $X(3915)$. Even if the mass is roughly plausible, the conventional $c\bar{c}$ structure of $X(3915)$ would likely encounter trouble in accommodating the product of the two-photon decay width and the branching ratio to $J/\psi \omega$, i.e. $\Gamma(\gamma\gamma)\times \mathcal{B}(X(3915)\rightarrow J/\psi \omega) = 61\pm 17\pm 8 $ eV for $J^P=0^+$ and $18\pm 5\pm 2 $ eV for $J^P=2^+$. As shown in Table \ref{gamma}, the two-photon decay width is $\Gamma_{\gamma\gamma}(2^3P_2)= 0.39$ keV in the MNR potential model. Based on the central value of the world-average width, $\Gamma(X(3915)\rightarrow J/\psi \omega)$ is expected to be $1.29$ MeV ($J^{PC}=2^+$) or $4.38$ MeV ($J^{PC}=0^+$) approximately, which is over ten times larger than that of the typical charmonium. On the other hand, the charmonium candidate $\chi_{c0}(2^3P_0)$ with $J^P=0^{++}$ may be slightly better in $\Gamma(\gamma\gamma)$, but worse in mass spectrum. The strong-decay inspection of the $X(3915)$ also disfavor the structure as $\chi_{c0}(2^3P_0)$ \cite{yyc2010,4350cc}. It has been interpreted as the multiquark and molecule state \cite{3915yyc,3915bugg,3915molecule}. So far, the nature of $X(3915)$ is still murky, and the charmonium system have seemed inefficient to provide a consensus of interpretation.þþ

\subsection{$X(3940)$ and $Z(3930)$}
\label{sec:4.3}

The $X(3940)$ was seen by the Belle Collaboration in the recoiling spectrum of $J/\psi$ in the $e^+e^-$ annihilation process \cite{X3940belle}. The mass and width of $X(3940)$ are $M=3942^{+7}_{-6}\pm6$ MeV, $\Gamma=37^{+26}_{-15}\pm8$ MeV \cite{pdg2010}. This state is not seen to decay into $D\bar{D}$ but does decay into $D\bar{D}^{\ast}$, suggesting that it has unnatural parity and can be a candidate for the $\eta_c(3^1S_0)$. The predicted mass is much higher by the potential models \cite{prd72, screen03, DEbert03, prd75}, including the present work. Taking the coupling channel effect into account \cite{newstates}, the induced spin-spin splitting between $3^3S_1$ and $3^1S_0$ will be increased and consequently lower down the mass of $3^1S_0$ state. We predicate the radiative transitions and two-photon decay width of $3^1S_0$ for testing the nature of $X(3940)$ as the charmonium candidate.

The Belle Collaboration has observed $Z(3930)$ state \cite{z3930belle} in two-photon fusion, $\gamma\gamma\rightarrow Z(3930)\rightarrow DD$. The world-average mass and full width are $M=3927.2\pm2.6$ MeV, $\Gamma=24\pm6$ MeV \cite{pdg2010}. And the theoretical prediction of mass within our framework is almost the same as the perturbative results, nearly 40 MeV higher than the observation. Due largely to the agreement with the detected angular distribution \cite{z3930belle,Z3930babar} for $J=2$ and to the plausible mass of a $2P$ $c\bar{c}$ state, $Z(3930)$ is now widely
accepted as $\chi_{c2}(3930)$ \cite{updatCC,status,pdg2010,prd32,z3930belle}. The value of the partial width $\Gamma(\gamma\gamma)\times\mathcal{B}(Z(3930)\rightarrow D\bar{D}) = 0.24\pm0.05\pm0.04$ keV was found in the Babar experiment \cite{Z3930babar}. Our prediction of $\Gamma(\gamma\gamma)$ is $0.39$ keV and $0.43$ keV within the MNR model and the NR model respectively. With the theoretical open-charm strong decay width $\Gamma(Z(3930)\rightarrow D\bar{D})=21.5$ MeV \cite{updatCC}, one can directly evaluate $\Gamma(\gamma\gamma)\approx 0.27\pm0.07$ keV. Our prediction of the two-photon width for $\chi_{c2}(2^3P_2)$ is reasonably consistent with this estimation, supporting $Z(3930)$ assigned as charmonium $2^3P_2$.

\subsection{$X(4160)$}
\label{sec:4.4}

The Belle Collaboration has reported the $X(4160)$ in the process $e^{+} e^{-}\rightarrow J/\psi D^{\ast}\bar{D}^{\ast}$ with a significance of 5.1 $\sigma$ \cite{4160exp}. Its mass and width are $M = 4156^{+25}_{-20}\pm 15 $ MeV and $\Gamma = 139 ^{+111}_{-61}\pm21$ MeV respectively. Since within this process $e^{+}e^{-}$ annihilate into one photon, the charge parity of $X(4160)$ should be $C=+$. The possible interpretation of its structure from the view of the production rate in $e^+ e^- \rightarrow J/\psi X(4160)$ favors the candidate $\chi_{c0}(3^3P_0)$ \cite{4160chao}, although the predicted mass differs around 30 MeV in the color screened potential model \cite{screen03}. In our calculated spectrum shown in Table \ref{cc}, our prediction of $\eta_c{(2^1D_2)}$ is $M(2^1D_2)=4164.9$ MeV in MNR model and $4173.7$ MeV in NR model. It is suggested that $X(4160)$ can be excellently reproduced by the charmonium spectrum as $\eta_c{(2^1D_2)}$ in our calculation framework. Investigating the strong decay properties of $X(4160)$, Ref. \cite{yyc2010} has indicated that the decay width of $\eta_c{(2^1D_2)}$ agrees well with the Belle data of $X(4160)$, which includes the absence of theoretical forbidden channel $\eta_c{(2^1D_2)}\rightarrow D\bar{D}$. Other possible candidates, i.e. $\chi_{c0}(3^3P_0)$, $\chi_{c1}(3^3P_1)$,$\eta_{c}(4^1S_0)$ are disfavored from this perspective. The charmonium structure of $X(4160)$ as $\eta_c{(2^1D_2)}$ can be further explored by electromagnetic transition rates, which have been predicted in this paper as well.

\subsection{$X(4350)$}
\label{sec:4.5}

The $X(4350)$ found in the invariant mass spectrum of $J/\psi \phi$ by the Belle Collaboration \cite{4350exp}, is of $M =4350^{+4.6}_{-5.1}\pm 0.7$ MeV and $\Gamma = 13.3^{+17.9}_{-9.1}\pm4.1$ MeV. Its assignment $J^{PC}$ could be $0^{++}$ or $2^{++}$. As a new charmonium-like state, so far, it has
been interpreted as $P$-wave charmonium state \cite{4350cc2010LX,4350cc}, the $D^{\ast}_s D^{\ast}_{s0}$ molecule \cite{4350molecule01,4350molecule02}, the $c\bar{c}s\bar{s}$ state \cite{4350tetra} and a mixture of scalar $c\bar{c}$ and $D^{\ast}_s\bar{D}^{\ast}_s$ \cite{4350mix} etc.. In particular,
Ref. \cite{4350cc2010LX} has presented the open-charm decay behaviors of the possible $c\bar{c}$ options, i.e. $\chi_{cJ}(3P)$ ($J=0,1,2$), using the quark pair creation (QPC) model. Only the decay properties of $\chi_{c2}(3^3P_2)$ are consistent with the existing experimental data on the $X(4350)$. Besides, the hidden charm decay $\Gamma(X(4350)\rightarrow J/\psi \phi)$ \cite{4350cc} via $D\bar{D}^{(\ast)}$ intermediate states favors the assignment
being $P$-wave charmonium as well.

The mass of $\chi_{c2}(3^3P_2)$ in our calculation is $4383.4$ MeV (NR) and $4352.4$ MeV (MNR). Consequently, the quantum number of the $X(4350)$ is reasonably favored as $J^{PC}= 2^{++}$ with the charmonium state $\chi_{c2}(3^3P_2)$ from the current exact mass prediction and strong decay behavior. To confirm this new member of charmonium family, more experimental data are required. We provide the corresponding radiative transition rates (see Table \ref{ccE1-2} for E1 channels and Table \ref{ccM1} for M1) and the two-photo decay width $\Gamma_{\gamma \gamma}( 3^3P_2) = 0.66$ keV for the further
testing.

\subsection{M1 transition of $J/\psi$ and $\psi^{\prime}$}
\label{sec:4.6}

A well-known problem is evident in the decay rate of $J/\psi \rightarrow \gamma \eta_c$, which is that the
 predicted rate in the nonrelativistic potential model is about a factor of 2-3 larger than experiment in the previous
 calculations\cite{prd72,SD2004}. Since this rate only involves the charm quark magnetic moment, this discrepancy leads
 to a surprise. When we adopt the wave functions generated from the full-potential Hamiltonian, this unexpected
 discrepancy seems can be reduced.

The radiative transition rates of $\psi^{\prime}$ from the previous perturbative calculation, both NR model and GI model, have showed serious disagreement with theory. Particularly, the predicted partial decay width of $\psi^{\prime}\rightarrow\gamma\eta_c$ was nearly one order of magnitude larger than the experimental observation \cite{pdg2010}. The lattice QCD calculation \cite{ccLQCD02} obtained $\psi^{\prime}\rightarrow\gamma\eta_c =
0.4\pm 0.8$ keV, while its uncertainty was apparently large. Recently, Li and Zhao has interpreted this discrepancy of potential model via taking the intermediate meson loop (IML) into account \cite{ZQ2011}. The IML contributions bring down the GI amplitude to $2.05^{+2.650}_{-1.75}$ keV, where the cut-off parameter played a crucial and sensitive role. Besides, the uncertainties from IML also due to the radiative couplings. We re-examined the NR model through the full-potential calculation, and found $\Gamma(\psi^{\prime}\rightarrow\gamma\eta_c) = 3.1$ keV (see Table \ref{ccM1}). In comparison with the perturbative results, the present approach may obtain a better wave function. For $\psi^{\prime}\rightarrow\gamma\eta_c^{\prime}$, our prediction is also in a reasonable agreement with the first measurement \cite{BES2011} of this decay channel by BES-III Collaboration very
recently.

\section{Remarks}
\label{sec:5}
By mean of a proper variational method, we calculate precisely the eigen value and eigen function of the Hamiltonian including the full interactions, which might maintain the subtleness of wave functions. We investigate the mass spectrum, electromagnetic transitions, leptonic decays and two-photon widths of the charmonium states. Comparing the results from the pure scalar and the scalar-vector mixing linear confinement, it is revealed that the mixture might be important for reproducing the mass spectrum and decay widths. Explicitly, about $22 \% $ vector component is obtained via
optimizing the potential parameters. The scalar-vector mixing model requires less linear slope, and more in the spin-splitting than the
pure scalar confinement. In our calculations, the M1 radiative transitions rates of $J/\psi$ and $\psi^{\prime}$ is reduced by the obtained wave functions actually, consisting with the experimental observations in order. The $S$-$D$ mixing are explored in the higher charmonium states. To the newly found charmonium-like states, we perform a reasonable inspection of the possible $c\bar{c}$ assignments. The $X(4160)$ and $X(4350)$ are found in the mass spectrum as $M(2^1D_2)= 4164.9$ MeV and $M(3^3P_2)= 4352.4$ MeV, which strongly favor the $J^{PC}=2^{-+}, 2^{++}$ assignments respectively. With the available experimental data, the $Z(3930)$ is favored as $\chi_{c2}(2^3P_2)$, while the $X(3872)$ and $X(3915)$ are disfavored by charmonium states. The corresponding radiative transitions, leptonic and two-photon decay widths of the $X(3940)$, $X(4160)$ and $X(4350)$ have been calculated for the further
experimental search and identifying the new members of charmonium family.

\begin{acknowledgements}
This work was supported in part by the National Natural Science Foundation of China (No.11175146, No.11047023) and the Fundamental Research Funds for the Central Universities (No.XDJK2012D005).
\end{acknowledgements}

\section*{Appendix: Wave functions and matrix elements for Gaussian basis expansion}

\setcounter{section}{1}
 \setcounter{equation}{0}
 \renewcommand{\theequation}{\Alph{section}.\arabic{equation}}

Since the state wave function $\psi_{lm}(\textbf{r})$ are expanded in the Gaussian basis
 space, the Hamiltonian matrix elements and the radial wave function at the origin, as well as the overlap integral involved in the transition matrix elements are explicitly analytic. We present these formulas used in the numerical calculation of the mass and decay width as follows.

\begin{equation}\label{1/r}
\langle\phi_{nlm}^{G}|\frac{1}{r}|\phi_{n^{\prime}lm}^{G}\rangle=
\frac{2}{\sqrt{\pi}}\frac{2^l l!}{(2l+1)!!}\sqrt{v_n+v_{n^{\prime}}}
\left( \frac{2\sqrt{v_n v_{n^{\prime}}}}{v_n+v_{n^{\prime}}}\right )^{l+\frac{3}{2}}
\end{equation}

\begin{equation}\label{r}
\langle\phi_{nlm}^{G}|r|\phi_{n^{\prime}lm}^{G}\rangle=
\frac{2^{l+1}}{\sqrt{\pi}}\frac{(l+1)!}{(2l+1)!!}\frac{1}{\sqrt{v_n+v_{n^{\prime}}}}\left(\frac{2\sqrt{v_n v_{n^{\prime}}}}{v_n+v_{n^{\prime}}}\right)^{l+\frac{3}{2}}
\end{equation}

\begin{equation}\label{er2}
\langle\phi_{nlm}^{G}\left|e^{-\sigma^2r^2}\right|\phi_{n^{\prime}lm}^{G}\rangle=
\left(\frac{2\sqrt{v_n v_{n^{\prime}}}}{v_n + v_{n^{\prime}}+ \sigma^2}\right )^{l+\frac{3}{2}}
\end{equation}

\begin{equation}\label{1/r3}
\langle\phi_{nlm}^{G}|\frac{1}{r^3}|\phi_{n^{\prime}lm}^{G}\rangle=
\frac{2^{l+1}}{\sqrt{\pi}}\frac{(l-1)!}{(2l+1)!!}\left(v_n+v_{n^{\prime}}\right)^{\frac{3}{2}}
\left(\frac{2\sqrt{v_nv_{n^{\prime}}}}{v_n+v_n^{\prime}}\right)^{l+\frac{3}{2}}
\end{equation}

\begin{equation}\label{RS}
R_{nS}(0)= \sum^{n_{max}}_{x=1}\frac{2\times2^{3/4}\sqrt{v_x^{3/2}}}{\pi^{1/4}}
\end{equation}

\begin{equation}\label{RD}
R^{\prime\prime}_{nD}(0)=\sum^{n_{max}}_{x=1}\frac{16\times2^{3/4}\sqrt{v_x^{7/2}}}{\sqrt{15}\pi^{1/4}}
\end{equation}

\begin{equation}\label{RP}
R^{\prime}_{nP}(0)=\sum^{n_{max}}_{x=1}\frac{4\times2^{3/4}\sqrt{v_x^{5/2}}}{\sqrt{3}\pi^{1/4}}
\end{equation}

\begin{eqnarray}\label{em1}
\left\langle \psi_f \left|r\right| \psi_i \right\rangle &=& \int_0^\infty\, \psi_f(\vec{r})\,r\,\psi_i(\vec{r})r^2\,dr
= \sum^{n_{max}}_{x=1}\sum^{n_{max}}_{y=1}C_{x L_i}\,C_{y L_f} \langle\phi_{yL_f}^{G}|r|\phi_{xL_i}^{G}\rangle  \\
\left\langle \psi_f| \psi_i \right\rangle &=& \int_0^\infty\, \psi_f(\vec{r})\,\psi_i(\vec{r})r^2\,dr
= \sum^{n_{max}}_{x=1}\sum^{n_{max}}_{y=1}C_{x L_i}\,C_{y L_f} \langle\phi_{yL_f}^{G}|\phi_{xL_i}^{G}\rangle
\end{eqnarray}
\begin{eqnarray}\label{em2}
\langle\phi_{yL_f}^{G}|r|\phi_{xL_i}^{G}\rangle &=& \frac{2^{\frac{5}{2}+L_f+L_i}\sqrt{v_y^{L_f+\frac{3}{2}}v_x^{L_i+\frac{3}{2}}}}{\sqrt{\pi }\sqrt{(2
L_f+1)!!(2L_i+1)!!}}(v_y+v_x)^{\frac{1}{2}(-L_f-L_i-4)}\Gamma\left[\frac{1}{2} (L_f+L_i+4)\right]  \\
\langle\phi_{yL_f}^{G}|\phi_{xL_i}^{G}\rangle &=& \frac{2^{\frac{5}{2}+L_f+L_i}\sqrt{v_y^{L_f+\frac{3}{2}}v_x^{L_i+\frac{3}{2}}}}{\sqrt{\pi }\sqrt{(2L_f+1)!!(2L_i+1)!!}}(v_y+v_x)^{\frac{1}{2}(-L_f-L_i-3)}\Gamma\left[\frac{1}{2} (L_f+L_i+3)\right]
\end{eqnarray}


\begin{thebibliography}{66}
%
%
\bibitem[1]{puzzles}
Brambilla N, et al. (Quarkonium Working Group) (2011) Heavy quarkonium: progress, puzzles, and opportunities. Eur. Phys. J. C71: 1534, arXiv:1010.5827 [hep-ph]
\bibitem[2]{babar2010}
Biassoni P (2010) Recent results in charmonium spectroscopy at B-factories. arXiv:1009.2627 [hep-ex]
\bibitem[3]{recharm}
Voloshin M (2008) Review charmonium. Prog. Part. Nucl. Phys. 61: 455-511
\bibitem[4]{QWG169}
Eichten E, Gottfried K, Kinoshita T, Lane K D, Yan T M (1976) Interplay of confinement and decay in the
spectrum of charmonium. Phys. Rev. Lett. 36: 500-504
\bibitem[5]{prd17}
Eichten E, Gottfried K, Kinoshita T, Lane K D, Yan T M (1978) Charmonium: The model. Phys. Rev. D 17: 3090-3117
\bibitem[6]{prd21}
Eichten E, Gottfried K, Kinoshita T, Lane K D, Yan T M (1980) Charmonium: Comparison with experiment. Phys. Rev. D 21: 203-233
\bibitem[7]{prd32}
Godfrey S, Isgur N (1985) Mesons in a relativized quark model with chromodynamics. Phys. Rev. D 32: 189-231
\bibitem[8]{lhqq}
Stanley D P, Robson D (1980) Nonperturbative potential model for light and heavy quark-antiquark systems. Phys. Rev. D 21: 3180-3196
\bibitem[9]{DEbert02}
Ebert D, Faustov R N, Galkin V O (2003) Properties of heavy quarkonia and bc mesons in the relativistic
quark model. Phys. Rev. D 67: 014027
\bibitem[10]{Bc}
Gershtein S S, Kiselev V V, Likhoded A K, Tkabladze A V (1995) $B_c$ spectroscopy. Phys. Rev. D 51: 3613-3627
\bibitem[11]{sop-heavy}
Fulcher L P (1991) Perturbative QCD, a universal QCD scale, long-range spin-orbit potential, and the properties of heavy quarkonia. Phys. Rev. D 44: 2079-2084
\bibitem[12]{prd60-Bc}
Fulcher L P (1999) Phenomenological predictions of the properties of the $B_c$ system. Phys. Rev. D 60: 074006
\bibitem[13]{prd50}
Fulcher L P (1994) Matrix representation of the nonlocal kinetic energy operator, the spinless salpeter equation and the cornell potential. Phys. Rev. D 50: 447-453
\bibitem[14]{prd49hqp}
Gupta S N, Johnson J M, Repko W W, Suchyta C J (1994) Heavy quarkonium potential model and the $^1P_1$ state of charmonium. Phys. Rev. D 49: 1551-1555
\bibitem[15]{prd53-Bc}
Gupta SN, Johnson JM (1996) $B_c$ spectroscopy in a quantum-chromodynamic potential model. Phys. Rev. D 53: 312-314
\bibitem[16]{mesonRelat}
Zeng J, Van Orden J W, Roberts W (1995) Heavy mesons in a relativistic model. Phys. Rev. D 52: 5229-5241
\bibitem[17]{cornell01}
Eichten E, Gottfried K, Kinoshita T, Lane K D, Yan T M (1978) Charmonium: The model. Phys. Rev. D 17: 3090-3117
\bibitem[18]{cornell02}
Eichten E, Gottfried K, Kinoshita T, Lane K D, Yan T M (1980) Erratum: Charmonium: The model. Phys. Rev.D 21: 313
\bibitem[19]{flux}
Buchmuller W (1982) Fine- and hyperfine structure of quarkonia. Phys. Lett. B 112:479-483
\bibitem[20]{DEbert03}
Ebert D, Faustov R N, Galkin V O (2000) Quark-antiquark potential with retardation and radiative contributions
and the heavy quarkonium mass spectra. Phys. Rev. D 62: 034014
\bibitem[21]{prd72}
Barnes T, Godfrey S, Swanson E S (2005) Higher charmonia. Phys. Rev. D 72: 054026
\bibitem[22]{screen79}
Li B Q, Chao K T (2009) Higher charmonia and X, Y, Z states with screened potential. Phys. Rev. D 79: 094004
\bibitem[23]{prd75}
Radford S F, Repko W W (2007) Potential model calculations and predictions for heavy quarkonium. Phys. Rev. D 75: 074031
\bibitem[24]{gem}
Hiyama E, Kino Y, Kamimura M (2003) Gaussian expansion method for few-body systems. Prog. Part. Nucl. Phys. 51: 223-307
\bibitem[25]{pdg2010}
Nakamura K, et al. (Particle Data Group) (2010) The review of particle physics. J. Phys. G 37: 075021, and 2011 partial update
for the 2012 edition
\bibitem[26]{prd37}
Kwong W, Mackenzie P B, Rosenfeld R, Rosner J L (1988) Quarkonium annihilation rates. Phys. Rev. D 37:3210-3215
\bibitem[27]{screen03}
Ding Y B, Chao K T, Qin D H (1993) Screened Q-Q potential and spectrum of heavy quarkonium. Chin. Phys. Lett. 10: 460-463
\bibitem[28]{DEbert2003twophoton}
Ebert D, Faustov R N, Galkin V O (2003) Two-photon decay rates of heavy quarkonia in the relativistic
quark model. Mod. Phys. Lett. A 18: 601-608, arXiv:0302044 [hep-ph]
\bibitem[29]{SD2004}
Barnes T (2004) Charmonium at BES and CLEO-c. arXiv:0406327 [hep-ph]
\bibitem[30]{Gupta1996}
Gupta S N, Johnson J M, Repko W W (1996) Relativistic two-photon and two-gluon decay rates of heavy
quarkonia. Phys. Rev. D 54: 2075-2080
\bibitem[31]{Kwong1988}
Kwong W, Rosner J L (1988) D-wave quarkonium levels of the $\Upsilon$ family. Phys. Rev. D 38:279-297
\bibitem[32]{QWG}
Brambilla N, et al. (Quarkonium Working Group) (2005) CERN Yellow Report. arXiv:0412158 [hep-ph]
\bibitem[33]{BES2011}
Wang L, et al. (BESIII Collaboration) (2011) Study of charnomium spectroscopy at BESIII. arXiv:1110.2560 [hep-ex]
\bibitem[34]{2Sto1S2009}
Mitchell R, et al. (CLEO Collaboration) (2009) $J/\psi$ and $\psi(2S)$ radiative decays to $\eta_c$. Phys. Rev. Lett. 102: 011801
\bibitem[35]{Rosner2001}
Rosner J L (2001) Charmless final states and S-D-wave mixing in the $\psi^{\prime\prime}$. Phys. Rev. D 64: 094002, arXiv:0105327 [hep-ph]
\bibitem[36]{4160chao}
Chao K T (2008) Interpretations for the X(4160) observed in the double charm production at B factories. Phys. Lett. B 661: 348-353, arXiv:0707.3982 [hep-ph]
\bibitem[37]{SDmix01}
Badalian A, Bakker B, Danilkin I (2009) The S-D mixing and di-electron widths of higher charmonium $\mathbf{1^{--}}$ states. Phys. Atom. Nucl. 72: 638-646, arXiv:0805.2291 [hep-ph]
\bibitem[38]{3872expt}
Choi S K, et al. (Belle Collaboration) (2003) Observation of a narrow charmonium-like state in exclusive $B^+ \to K^+ \pi^+ \pi^- J/\psi$ decays. Phys. Rev. Lett. 91: 262001, arXiv:0309032 [hep-ex]
\bibitem[39]{3872jpc}
CDF Collaboration (2007) Analysis of the Quantum Numbers $J^{PC}$ of the X(3872) Particle. Phys. Rev. Lett. 98: 132002, arXiv:0612053 [hep-ex]
\bibitem[40]{3872gamma}
Aubert B, et al. (BABAR Collaboration) (2009) Evidence for X(3872) $\to \psi(2S) \gamma$ in $B^{\pm} \to$ X(3872) $K^{\pm}$ decays, and a study of $B \to c\bar{c} \gamma K$. Phys. Rev. Lett. 102: 132001, arXiv:0809.0042 [hep-ex]
\bibitem[41]{3872belle1++}
Abe K, et al. (Belle Collaboration) (2005) Evidence for X(3872)$\to \gamma J/\psi$ and the sub-threshold decay X(3872)$\to \omega J/\psi$. arXiv:0505037 [hep-ex]
\bibitem[42]{3872belle2-+}
del Amo Sanchez P, et al (BABAR Collaboration) (2010) Evidence for the decay X(3872)$\to J/\psi\omega$. Phys. Rev. D 82: 011101
\bibitem[43]{3915expt2010}
Uehara S, et al. (Belle Collaboration) (2010) Observation of a charmonium-like enhancement in the $\gamma \gamma \to \omega J/\psi$ process. Phys. Rev. Lett. 104: 092001, arXiv: 0912.4451 [hep-ex]
\bibitem[44]{3940Ybelle}
Choi S K, et al. (Belle Collaboration) (2005) Observation of a near-threshold $\omega-J/\psi$ mass enhancement in exclusive $B \to K \omega J/\psi$ decays. Phys. Rev. Lett. 94: 182002, arXiv:0408126 [hep-ex]
\bibitem[45]{3490Ybabar2008}
Aubert B, et al. (BABAR Collaboration) (2008) Observation of Y(3940) $\to J/\psi \omega$ in $B \to J/\psi \omega K$ at BABAR. Phys. Rev. Lett. 101: 082001, arXiv:0711.2047 [hep-ex]
\bibitem[46]{prd69}
Eichten E J, Lane K, Quigg C (2004) Charmonium levels near threshold and the narrow state X(3872)$\to \pi^{+}\pi^{-}J/\psi$. Phys. Rev. D 69: 094019
\bibitem[47]{yyc2010}
Yang Y C, Xia Z R, Ping J L (2010) Are the X(4160) and X(3915) charmonium states? Phys. Rev. D 81: 094003
\bibitem[48]{4350cc}
Zhao Z J, Pan D M (2011) Estimating strong decays of X(3915) and X(4350). arXiv:1104.1838 [hep-ph]
\bibitem[49]{3915yyc}
Yang Y C, Ping J L (2010) Dynamical study of the $X$(3915) as a molecular $D^*\bar{D^*}$ state in a quark model. Phys. Rev. D 81: 114025, arXiv:1004.2444 [hep-ph]
\bibitem[50]{3915bugg}
Bugg D V (2011) Explanation for Y(4140) and X(3915). arXiv:1103.5363 [hep-ph]
\bibitem[51]{3915molecule}
Branz T, Gutsche T, Lyubovitskij V E (2009) Hadronic molecule structure of the Y(3940) and Y(4140). Phys. Rev. D 80: 054019, arXiv:0903.5424 [hep-ph]
\bibitem[52]{X3940belle}
Abe K, et al. (Belle Collaboration) (2007) Observation of a new charmonium state in double charmonium production in $e^{+} e^{-}$ annihilation at $\sqrt{s}\sim$ 10.6 GeV. Phys. Rev. Lett. 98: 082001, arXiv:0507019 [hep-ex]
\bibitem[53]{newstates}
Eichten E J, Lane K, Quigg C (2006) New states above charm threshold. Phys. Rev. D 73: 014014
\bibitem[54]{z3930belle}
Uehara S, et al. (Belle Collaboration) (2006) Observation of a $\chi^{\prime}_{c2}$ candidate in $\gamma \gamma \to D\bar{D}$ production at Belle. Phys. Rev. Lett. 96: 082003, arXiv:0512035 [hep-ex]
\bibitem[55]{Z3930babar}
Aubert B, et al. (BABAR Collaboration) (2010) Observation of the $\chi_{c2}$(2P) meson in the reaction $\gamma \gamma \to D\bar{D}$ at BABAR. Phys. Rev. D
81: 092003, arXiv:1002.0281 [hep-ex]
\bibitem[56]{updatCC}
Barnes T (2010) Update on charmonium theory. arXiv:1003.2644 [hep-ph]
\bibitem[57]{status}
Swanson E S (2006) The new heavy mesons: A status report. Phys. Rep. 429: 243-305
\bibitem[58]{4160exp}
Pakhlov P, et al. (Belle Collaboration) (2008) Search for new charmonium states in the processes $e^{+} e^{-} \to J/\psi D^{*}D^{*}$ at $\sqrt{s}\sim$ 10.6 GeV. Phys. Rev. Lett. 100: 202001, arXiv:0708.3812 [hep-ex]
\bibitem[59]{4350exp}
Shen C P, et al. (Belle Collaboration) (2010) Evidence for a new resonance and search for the Y(4140) in $\gamma \gamma \to \varphi J/\psi$. Phys. Rev. Lett. 104: 112004, arXiv:0912.2383 [hep-ex]
\bibitem[60]{4350cc2010LX}
Liu X, Luo Z G, Sun Z F (2010) X(3915) and X(4350) as new members in the P-wave charmonium family. Phys. Rev. Lett. 104: 122001
\bibitem[61]{4350molecule01}
Zhang J R, Huang M Q (2010) $\{Q\bar{s}\}\{\bar{Q}^{(\prime)}s\}$ molecular states in QCD sum rules. Commun. Theor. Phys. 54: 1075-1090, arXiv:0905.4672 [hep-ph]
\bibitem[62]{4350molecule02}
Ma Y L (2010) Estimates for X(4350) decays from the effective lagrangian approach. Phys. Rev. D 82: 015013, arXiv:1006.1276 [hep-ph]
\bibitem[63]{4350tetra}
Stancu F (2010) Can Y(4140) be a $c\bar{c}s\bar{s}$ tetraquark? J. Phys. G 37: 075017
\bibitem[64]{4350mix}
Wang Z G (2010) Analysis of the X(4350) as a scalar $\bar{c}c$ and ${D}_s^\ast {\bar {D}}_s^\ast$ mixing state with QCD sum rules. Phys. Lett. B 690: 403, arXiv:0912.4626 [hep-ph]
\bibitem[65]{ccLQCD02}
Dudek J J, Edwards R G, Thomas C E (2009) Exotic and excited-state radiative transitions in charmonium from lattice QCD. Phys. Rev. D 79: 094504
\bibitem[66]{ZQ2011}
Li G, Zhao Q (2011) Revisit the radiative decays of $J/\psi$ and $\psi^{\prime} \to \gamma \eta_c (\gamma \eta_c^{\prime})$. Phys. Rev. D 84: 074005, arXiv:1107.2037 [hep-ph]
\bibitem{chen} Chen H, Ping R G (2009) Charmonium rescattering effects in $\psi^{\prime}\to\gamma\eta_{c}$ decay. Eur. Phys. J. A 42: 237
\end{thebibliography}


\end{document}